\documentclass[10pt]{iopart}
\usepackage{graphicx}
\pdfoutput=1
\usepackage{xcolor}
\usepackage{iopams} 

\eqnobysec
\newcommand{\beq}{\begin{equation}}
\newcommand{\beqa}{\begin{eqnarray}}
\newcommand{\eeq}{\end{equation}}
\newcommand{\eeqa}{\end{eqnarray}}

\renewcommand{\c}{{\cal C}}

\newcommand{\s}{{\sigma}}

\newcommand{\w}{{\bar w}}

\newcommand{\E}{{\bf e}}
\newcommand{\x}{{\bf x}}
\newcommand{\y}{{\bf y}}
\def\sign{\mathop{\rm sign}}

\begin{document}

\title{Rates for irreversible Gibbsian Ising models}
\date{\today}
\author{Claude Godr\`eche}
\address{
Institut de Physique Th\'eorique, CEA Saclay and CNRS\\
91191 Gif-sur-Yvette cedex, France}
\begin{abstract}
Dynamics under which a system of Ising spins relaxes to a stationary state  with Bolzmann-Gibbs measure and which do not fulfil the condition of detailed balance are irreversible and asymmetric.
We revisit the problem of the determination of rates yielding such a stationary state for models with single-spin flip dynamics.
We add some supplementary material to this study and confirm that Gibbsian irreversible Ising models exist for one and two-dimensional lattices but not for the three-dimensional cubic lattice.
We also analyze asymmetric Gibbsian dynamics in the limit of infinite temperature.
We finally revisit the case of a linear chain of spins under asymmetric conserved dynamics.
\end{abstract}

\maketitle

\section{Introduction}

The one-dimensional Glauber-Ising model was probably the first example of a strongly interacting system with soluble dynamics showing how a system relaxes to equilibrium~\cite{glau}.
The preliminary question to solve was the choice of rates ensuring that a ferromagnetic chain of Ising spins relaxes towards equilibrium under single spin flip dynamics.
The question is settled by requiring the rates at which spins flip to fulfil the condition of detailed balance with respect to the Hamiltonian defining the model~\cite{glau}.

Conversely, one may ask whether it is still possible for the system to reach a stationary state with the same Boltzmann-Gibbs measure by an appropriate choice of rates if one relinquishes this constraint, i.e., if only global balance is imposed.
The dynamics now becomes generically irreversible and asymmetric: the flipping spin is not equally influenced by its neighbours.
This question was addressed some time ago by K\"{u}nsch~\cite{kun}, who exhibited examples of such a dynamics, in the particular case where it is totally asymmetric, in one and two dimensions.
The problem was thoroughly revisited in ref.~\cite{gb2009}, with the following conclusions.

The study made in~\cite{gb2009} shows that irreversible Gibbsian Ising models exist for one and two-dimensional lattices but not for the three-dimensional cubic lattice.
More precisely, imposing the up-down spin symmetry, the rate function yielding an irreversible Gibbsian stationary state for the linear chain depends on 3 arbitrary parameters.
In two dimensions, the number of arbitrary parameters is respectively equal to 10 for the square lattice, and to 35 for the triangular lattice.
Yet, for the totally asymmetric dynamics where only half of the spins have an influence on the flipping spin, the rate function is {\it unique}, up to a time scale, for these three geometries.
In contrast, for the cubic lattice no such rate function is found, i.e., global balance enforces detailed balance (see Table~\ref{tab:result}).

The aim of the present work is to add some supplementary material to the same study, making the method used more easy to grasp and illustrating its outcomes on more examples.
In particular we give a fuller account of the method, which relies on linear algebra coupled to the properties of the system under translation invariance, in order to make clearer its generality.
We come back on the interplay between coordination and dimension.
We explain, on the example of the linear chain, the constraints imposed by the positivity of the rates.
We shall also be concerned in restating the statement made in~\cite{gb2009}, that no such rates do exist for the case of the three-dimensional cubic lattice, as recalled above.
To this end we shall give some more details on the analysis in order to substantiate its conclusion.
We shall then give a critical reading of a recent paper~\cite{oliv}, where it is claimed that there exist irreversible Gibbsian dynamics for the cubic lattice, in contradiction with the study made in~\cite{gb2009}, and shall dismiss its conclusion on this issue.
We shall finally consider the case of Gibbsian asymmetric dynamics at infinite temperature.
An appendix is devoted to the study of conserved Gibbsian asymmetric dynamics for the linear chain.\footnote[1]{The author of ref.~\cite{oliv} recently issued an Erratum~\cite{erratum} where he corrects the claims which lead him to the incorrect prediction mentioned in the Introduction. 
We nevertheless kept the text of section~\ref{sec:oliv} unchanged because the analysis presented there provides an interesting illustration, on the example chosen in ref.~\cite{oliv}, of the fact that global balance enforces detailed balance for the cubic lattice, and explains the mechanism by which this occurs.}
\section{Starting point}

Let us consider $N$ Ising spins on a regular lattice of coordination $z$, in $D$ dimensions, with periodic boundary conditions.
The energy (Hamiltonian) of a configuration $\c=\{\s_{1},\ldots,\s_{n},\ldots,\s_{N}\}$ reads
\beq\label{hamilt}
E({\cal C})=-J\sum_{n,j}\s_n\s_{j},
\eeq
where $n$ and $j$ are nearest neighbours.

\subsection{Master equation}

The dynamics consists in flipping a spin, chosen at random, say spin $n$, with a rate 
$w(\c_n|\c)$, corresponding to the transition between configurations $\c$ and 
$\c_n=\{\s_{1},\ldots,-\s_{n},\ldots,\s_{N}\}$.
At stationarity, the master equation expresses that losses are equal to gains, and reads
\beq\label{master0}
P(\c)\sum_{n}w(\c_n|\c)=\sum_{n}w(\c|\c_n)P(\c_n).
\eeq
We want to find the rate function $w(\c_n|\c)$ satisfying this equation when $P(\c)$ is the Boltzmann-Gibbs distribution associated to the Hamiltonian~(\ref{hamilt}),
\beq
P(\c)\propto\e^{-E(\c)/T}.
\eeq
After division of both sides by the weight $P(\cal C)$,
eq.~(\ref{master0}) can be rewritten as
\beq\label{master}
\sum_{n}\left( w(\c_n|\c)-w(\c|\c_n)\e^{-\Delta E/T}\right)=0,
\eeq
where the change in energy due to the flip reads
\beq\label{del}
\Delta E=E(\c_n)-E(\c)=2J\,\s_n h_n,
\eeq
and $h_n$ is the local field $h_n=\sum_j \s_j$, due to the $z$ neighbours $\{\s_{j}\}$.
We choose a rate function only depending on the local configuration 
$\{\s_n;\{\s_{j}\}\}$ of the central spin $\s_n$ and of its neighbours, and simplify the notation accordingly,
\beq
w(\c_n|\c)=w(\s_n;\{\s_{j}\}).
\eeq
Thus, denoting the balance term by
\beq\label{balance}
B(\s_n;\{\s_{j}\})= w(\s_n;\{\s_{j}\})-w(-\s_n;\{\s_{j}\})\e^{-2K \s_n h_n},
\eeq
where $K=J/T$, the balance equation~(\ref{master}) becomes
\beq\label{mast}
\sum_{n} B(\s_n;\{\s_{j}\})=0.
\eeq
This equation can be satisfied either term by term, which gives the detailed balance condition on the rate function,
\beq\label{db}
w(\s_n;\{\s_{j}\})=w(-\s_n;\{\s_{j}\})\e^{-2K \s_n h_n}.
\eeq
or as a whole, which is the global balance condition.

\subsection{Representation of the rate function on a basis of spin operators}
The number of values taken by the rate function is equal to the number of local configurations
$\{\s_n;\{\s_{j}\})\}$ of the central spin and of its neighbours.
There are $2^{z+1}$ such configurations, i.e., 8 for the chain, 32 for the square lattice, and 128 for the two-dimensional triangular lattice or for the cubic lattice.
We hereafter consider the simpler case where we have up-down spin symmetry: 
\beq\label{symm}
w(\s_n;\{\s_j\})=w(-\s_n;\{-\s_j\}).
\eeq
The number of possible values of the rate function is therefore halved and is equal to the number of different environments of the central spin $\s_n$, i.e., of configurations $\{\s_j\}$ of its neighbours.
There are $2^z$ such configurations, labelled by the index $\alpha$,
i.e., 4 for the chain, 16 for the square lattice, and 64 for the triangular lattice or for the cubic lattice.
We denote the $2^z$ rates with $\s_n=+1$ by $w_{\alpha}$ and the other $2^{z}$ rates, corresponding to $\s_n=-1$, by $\bar w_{\alpha}$:
\beq\label{wal}
w_{\alpha}=w(\s_n=+1;\{\s_j\}_\alpha),\qquad \bar w_\alpha=w(\s_n=-1;\{\s_j\}_\alpha).
\eeq
The latter are obtained from the former by the spin symmetry relation~(\ref{symm}), 
yielding
\beq\label{wsymm}
\bar w_\alpha=w_{2^z+1-\alpha}, 
\eeq
(see Tables~\ref{tab:1d} and~\ref{tab:2d}).
For instance, for the linear chain, the rates to be determined are
\beqa\label{rate1D}
w_1=w(+;++),\quad w_2=w(+;+-),
\nonumber\\
w_3=w(+;-+),\quad w_4=w(+;--).
\eeqa

\begin{table}[ht]
\caption{List of local configurations and corresponding values of the rate function for the one-dimensional chain.
There are 4 possible rates $w_{\alpha}$, with $\s_n=+1$, corresponding to the 4 possible configurations $\{\s_j\}$, labelled by $\alpha$, of the two neighbours of the central spin, taken in the order: left, right.
The 4 remaining rates $\w_\alpha$, with $\s_n=-1$, are deduced from the former, due to the spin symmetry (see~(\ref{wsymm})).}
\label{tab:1d}
\begin{center}
\begin{tabular}{|c||c|c||c|c|}
\hline
$\alpha$&$\s_n;\{\s_j\}$&$w_\alpha$&$\s_n;\{\s_j\}$&$\w_\alpha$\\
\hline
$1$&$+;++$&$w_{1}$&$-;++$&$\w_1=w_4$\\
$2$&$+;{+-}$&$w_2$&$-;+-$&$\w_2=w_3$\\
$3$&$+;{-+}$&$w_3$&$-;-+$&$\w_3=w_2$\\
$4$&$+;{--}$&$w_4$&$-;--$&$\w_4=w_1$\\
\hline
\end{tabular}
\end{center}
\end{table}
\begin{table}[ht]
\caption{List of local configurations and corresponding values of the rate function for the 2D square lattice.
There are 16 possible rates $w_{\alpha}$, with $\s_n=+1$, corresponding to the 16 possible configurations $\{\s_j\}$, labelled by $\alpha$, of the four neighbours of the central spin, taken in the order: east, north, west, south.
The 16 remaining rates $\w_\alpha$, with $\s_n=-1$, are deduced from the former, due to the spin symmetry (see~(\ref{wsymm})).}
\label{tab:2d}
\begin{center}
\begin{tabular}{|c||c|c||c|c|}
\hline
$\alpha$&$\s_n;\{\s_j\}$&$w_\alpha$&$\s_n;\{\s_j\}$&$\w_\alpha$\\
\hline
$1$&$+;{++++}$&$w_{1}$&$-;++++$&$\w_1=w_{16}$\\
$2$&$+;{+++-}$&$w_2$&$-;+++-$&$\w_2=w_{15}$\\
$3$&$+;{++-+}$&$w_3$&$-;{++-+}$&$\w_3=w_{14}$\\
$4$&$+;{++--}$&$w_4$&$-;++--$&$\w_4=w_{13}$\\
$5$&$+;+-++$&$w_{5}$&$-;+-++$&$\w_5=w_{12}$\\
$6$&$+;+-+-$&$w_{6}$&$-;+-+-$&$\w_6=w_{11}$\\
$7$&$+;+--+$&$w_{7}$&$-;+--+$&$\w_7=w_{10}$\\
$8$&$+;+---$&$w_{8}$&$-;+---$&$\w_8=w_{9}$\\
$9$&$+;-+++$&$w_{9}$&$-;-+++$&$\w_9=w_{8}$\\
$10$&$+;-++-$&$w_{10}$&$-;-++-$&$\w_{10}=w_{7}$\\
$11$&$+;-+-+$&$w_{11}$&$-;-+-+$&$\w_{11}=w_{6}$\\
$12$&$+;-+--$&$w_{12}$&$-;-+--$&$\w_{12}=w_{5}$\\
$13$&$+;--++$&$w_{13}$&$-;--++$&$\w_{13}=w_{4}$\\
$14$&$+;--+-$&$w_{14}$&$-;--+-$&$\w_{14}=w_{3}$\\
$15$&$+;---+$&$w_{15}$&$-;---+$&$\w_{15}=w_{2}$\\
$16$&$+;----$&$w_{16}$&$-;----$&$\w_{16}=w_{1}$\\
\hline
\end{tabular}
\end{center}
\end{table}

When the spin symmetry is not imposed, the rate function depends on the values taken by the $z+1$ spins 
$\s_n$ and $\{\s_j\}$, and can be decomposed on a basis of $2^{z+1}$ spin operators made of $0, 1,\ldots, z+1$ spins.
For instance, for the linear chain, these operators are:
$\{1,\s_{n-1},\s_{n},\s_{n+1},\s_{n}\s_{n+1},\s_{n-1}\s_{n},$ $\s_{n-1}\s_{n+1},\s_{n-1}\s_{n}\s_{n+1}\}$.
In the present situation where spin symmetry holds, this decomposition can be restricted to $2^{z}$ even spin operators $O_i$, i.e.,
\beq\label{basis}
w(\s_n;\{\s_{j}\})=\sum_{i=0}^{2^{z}-1} c_i\, O_i,
\eeq
with $O_0=1$.
The knowledge of the $2^{z}$ coefficients $c_i$ is equivalent to the knowledge of the $2^{z}$ rates $w_\alpha$.
The coefficient $c_0$ fixes the scale of time.

For instance, for the linear chain,
\beq\label{oper1D}
w(\s_n;\{\s_{j}\})=c_0+c_1\,\s_{n}\s_{n+1}+c_2\,\s_{n-1}\s_{n}+c_3\,\s_{n-1}\s_{n+1},
\eeq
i.e.,
\beq\label{eq:oper1D}
O_1=\s_{n}\s_{n+1},\quad O_2=\s_{n-1}\s_{n}, \quad O_3=\s_{n-1}\s_{n+1}.
\eeq

For the square lattice, we use the following notations.
The central spin $\s_n$ being located at $\x_n$, we denote by $\s_{j_a}$ (resp. $\s_{j_{\underline{a}}}$) the neighbouring spins located at $\x_n+\E_a$ (resp. $\x_n-\E_a$), where $\E_a$ ($a=1,2$) are the unit vectors spanning the square lattice.
Thus $\s_{j_1}$, $\s_{j_2}$, $\s_{j_{\underline 1}}$ and $\s_{j_{\underline 2}}$ are the east, north, west and south spins, respectively.
The list of even operators is given in Table~\ref{tab:op2D}.
\begin{table}[ht]
\caption{List of the even operators made of the five spins ($\s_n;\{\s_j\}$).
The central spin $\s_n$ being located at $\x_n$, $\s_{j_a}$ (resp. $\s_{j_{\underline{a}}}$) are the neighbouring spins located at $\x_n+\E_a$ (resp. $\x_n-\E_a$), where $\E_a$ ($a=1,2$) are the unit vectors spanning the square lattice.}
\label{tab:op2D}
\begin{center}
\begin{tabular}{|l|l|}
\hline
$i$&$O_i$\\
\hline
$1$&$\s_n\s_{j_1}\s_{j_2}\s_{j_{\underline 1}}$\\
$2$&$\s_n\s_{j_1}\s_{j_2}\s_{j_{\underline 2}}$\\
$3$&$\s_n\s_{j_1}\s_{j_{\underline 1}}\s_{j_{\underline 2}}$\\
$4$&$\s_n\s_{j_2}\s_{j_{\underline 1}}\s_{j_{\underline 2}}$\\
$5$&$\s_{j_1}\s_{j_2}\s_{j_{\underline 1}}\s_{j_{\underline 2}}$\\
$6$&$\s_n\s_{j_1}$\\
$7$&$\s_n\s_{j_2}$\\
$8$&$\s_n\s_{j_{\underline 2}}$\\
$9$&$\s_n\s_{j_{\underline 1}}$\\
$10$&$\s_{j_1}\s_{j_2}$\\
$11$&$\s_{j_2}\s_{j_{\underline 1}}$\\
$12$&$\s_{j_1}\s_{j_{\underline 1}}$\\
$13$&$\s_{j_{\underline 1}}\s_{j_{\underline 2}}$\\
$14$&$\s_{j_1}\s_{j_{\underline 2}}$\\
$15$&$\s_{j_2}\s_{j_{\underline 2}}$\\
\hline
\end{tabular}
\end{center}
\end{table}

In order to determine the rate function satisfying the global balance condition~(\ref{mast}), or the detailed balance condition~(\ref{db}), we can proceed in either of two ways.
The first one consists in finding the constraints on the rates $\{w_\alpha\}$, from which constraints on the coefficients $\{c_i\}$ ensue.
The second one consists in finding the constraints on the coefficients $\{c_i\}$, from which constraints on the rates $\{w_\alpha\}$ ensue.
These two ways are strictly equivalent because the rates $\{w_\alpha\}$ are linear combinations of the coefficients $\{c_i\}$, and both are equivalent representations of the rate function $w(\s_n;\{\s_j\})$.

We emphasize this equivalence as follows.
Defining the indicator variables
\beq
I_\alpha=I(\s_n=+1;\{\s_j\}_\alpha),\quad \bar I_\alpha= I(\s_n=-1;\{\s_j\}_\alpha),
\eeq
we have, using the notation~(\ref{wal}),
\beq
w(\s_n;\{\s_{j}\})=\sum_{\alpha=1}^{{2^{z}}}\left(I_\alpha\,w_\alpha 
+\bar I_\alpha\,\bar w_\alpha\right).
\eeq
Using the spin symmetry relation~(\ref{wsymm}), we can rewrite the expression above as
\beq\label{eq:dec}
w(\s_n;\{\s_{j}\})=\sum_{\alpha=1}^{{2^{z}}} w_\alpha\left(I_\alpha+\bar I_{2^z+1-\alpha} 
\right),
\eeq
where the two indicator variables in the bracket correspond to two opposite configurations.
These indicator variables can be decomposed on the complete basis of $2^{z+1}$ spin operators
made of $0, 1,\ldots, z+1$ spins, however their sum only contains even operators $O_i$.
For instance, for the linear chain, 
\beqa\label{eq:decomp}
I_\alpha+\bar I_{2^z+1-\alpha}&=&
\frac{1+\s_n}{2}\frac{1+\sign(\s_{n-1})_{\alpha}\,\s_{n-1}}{2}
\frac{1+\sign(\s_{n+2})_{\alpha}\,\s_{n+1}}{2}
\nonumber\\
&+&
\frac{1-\s_n}{2}\frac{1-\sign(\s_{n-1})_{\alpha}\,\s_{n-1}}{2}
\frac{1-\sign(\s_{n+2})_{\alpha}\,\s_{n+1}}{2}
\nonumber\\
&=&\frac{1}{4}(1+
\sign(\s_{n+1})_{\alpha}\,\s_{n}\s_{n+1}+\sign(\s_{n-1})_{\alpha}\,\s_{n-1}\s_{n}
\nonumber\\
&+&\sign(\s_{n-1}\s_{n+1})_{\alpha}\,\s_{n-1}\s_{n+1})
\nonumber\\
&=&\sum_{i=0}^{2^z-1}a_{i,\alpha}O_i,
\eeqa
where we have introduced the matrix of signs (up to the constant $1/2^z$)
\beq
A=(a_{i,\alpha})= \frac{1}{2^z}O_i(\s_n=+1;\{\s_j\}_{\alpha}).
\eeq
(The matrix $(a_{i,\alpha})$ for the linear chain is given in the Appendix.)
One can thus rewrite (\ref{eq:dec}) as
\beq
w(\s_n;\{\s_{j}\})=\sum_{\alpha=1}^{{2^{z}}} w_\alpha\sum_{i=0}^{2^z-1}a_{i,\alpha}O_i.
\eeq
Identifying~(\ref{eq:dec}) with~(\ref{basis}) we obtain
\beq
c_i=\sum_{\alpha=1}^{{2^{z}}} a_{i,\alpha}w_\alpha.
\eeq
The inverse relation reads
\beq
w_\alpha=\sum_{i=0}^{2^{z}-1}c_i\,O_i(\s_n=+1;\{\s_j\}_{\alpha}).
\eeq
In other words
$A^2=I/2^z$, where $I$ is the unit matrix.

\subsection{Balance term}
Starting from (\ref{basis}), then using the identity $\e^{-a\, \s}=\cosh a-\sinh a\s$,
we can decompose the balance term on the basis of spin operators as
\beq\label{ee}
B(\s_n;\{\s_j\})=\sum_{i=0}^{2^{z}-1} E_i(\{c_i\})\, O_i,
\eeq
where the coefficients $E_i(\{c_i\})$ are linear combinations of the $c_i$, with coefficients depending on temperature through hyperbolic functions of $2K$. (See the Appendix for an illustration on the example of the linear chain.)

Let us now define
\beq\label{balph}
B_\alpha=w_\alpha-\w_\alpha\e^{-\beta \Delta E_\alpha},
\quad \bar B_\alpha=\w_\alpha-w_\alpha\e^{\beta \Delta E_\alpha},
\eeq
where $\Delta E_\alpha$ is the change of energy associated to the rate $w_\alpha$.
Thanks to the symmetry relation~(\ref{wsymm}) we have
\beq\label{eq:bb}
\bar B_\alpha=B_{2^z+1-\alpha}.
\eeq
For instance, for the linear chain, (\ref{eq:bb}) reads
\beqa
\bar B_1=\w_1-\e^{4K}w_1=w_4-\e^{4K}\w_4=B_4,
\nonumber\\
\bar B_2=\w_2-w_2=w_3-\w_3=B_3,
\eeqa
and so on.
Proceeding as for the rate function $w(\s_n;\{\s_{j}\})$, we can decompose the balance term $B(\s_n;\{\s_{j}\})$ as
\beq\label{eq:ff}
B(\s_n;\{\s_j\})=\sum_{i=0}^{2^{z}-1} F_i(\{B_\alpha\})\, O_i,
\eeq
where the coefficients $F_i(\{B_\alpha\})$ are linear combinations of the $B_\alpha$, 
\beq\label{Fi}
F_i=\sum_{\alpha=1}^{{2^{z}}}a_{i,\alpha}B_\alpha.
\eeq
(See the Appendix for the example of the linear chain.)
The sets $E_i$ and $F_i$ provide two equivalent representations of the linear decomposition of the balance term on the basis of spin operators, the former expressed in terms of the coefficients $c_i$, the latter in terms of the values $B_\alpha$ taken by the balance term.

\section{Detailed balance}
\label{detbal}

We start with the simple case of detailed balance, $B(\s_n;\{\s_j\})=0$, as a preparation for the sequel.
This equation is satisfied by imposing $E_i=F_i=0$ for all $i$.

\subsection{Constraints on the rates or on the coefficients}

The condition of detailed balance on the rate function~(\ref{db}) implies $2^z$ relations $\{B_\alpha=0\}$, or equivalently, $2^z$ relations $\{F_i=0\}$.
However, thanks to the symmetry relation~(\ref{eq:bb}) these relations are redundant and only half of them remain.
We thus get $2^{z-1}$ relations between pairs of rates:
\beq\label{eq:db}
w_\alpha=\w_\alpha \e^{-\beta \Delta E_\alpha},\qquad (\alpha=1,\ldots,2^{z-1}),
\eeq
In return, using the spin operator representation~(\ref{basis}) in these relations, the coefficients $c_i$ are found to obey $2^{z-1}$ linear constraints.

One can also proceed in reverse order, determining first the constraints on the coefficients $c_i$, then deducing those for the rates from the former. 
Expressing that $B(\s_n;\{\s_j\})$ vanishes identically, and using~(\ref{ee}) yields an homogeneous system of $2^z$ linear equations 
$\{E_i=0\}$, which are
not all independent.
The rank of this system is necessarily equal to the rank of the system $\{F_i=0\}$, i.e., to the number of relations between pairs of rates mentioned above, namely $2^{z-1}$.

\subsection{Examples}

We illustrate the previous considerations by the following examples.

For the linear chain, the relations~(\ref{eq:db}) are $B_1=B_2=0$, i.e.,
\beqa\label{res1D}
w(+;++)=\e^{-4K}w(-;++),\nonumber\\
w(+;+-)=w(-;+-).\label{constr}
\eeqa
The constraints on the coefficients are either deduced from~(\ref{res1D}) or obtained from the solution of the 4 equations 
$\{E_i=0\}$ (see~\ref{app:1D}):
\beq\label{lin}
c_1+c_2+\gamma(c_0+c_3)=0,\quad c_1=c_2,
\eeq
where 
\beq
\gamma=\tanh 2 K.
\eeq
The space of independent rates, or independent coefficients, has dimension 2.
We thus find the most general rate function obeying detailed balance
\beq\label{glau}
w(\s_n;\{\s_{j}\})=
\frac{\alpha}{2}\big(1+\delta\s_{n-1}\s_{n+1}
-\frac{\gamma}{2}(1+\delta)\s_{n}(\s_{n-1}+\s_{n+1})\big),
\eeq
where $\alpha$ and $\delta$ are the free parameters, recovering a result due to Glauber, written here with his notations~\cite{glau}.

On the square lattice, there are 8 constraints $B_1=B_2=\cdots= B_8=0$, on the 16 rates $w_\alpha$:
\beqa\label{res2D}
w_{1}=\e^{-8 K}\, \w_{1},\nonumber \\
w_{\alpha}=\e^{-4 K}\, \w_{\alpha},\quad &(&\alpha=2, 3, 5),\nonumber \\
w_{\alpha}=\w_{\alpha}.\quad &(&\alpha=4, 6, 7),\nonumber\\
w_{8}=\e^{4 K}\,\w_{8}.
\eeqa
We do not write down the corresponding 8 relations between the $c_i$ because we will not use them in the sequel.

\subsection{Symmetric rates}
\label{sec:sym_rates}
Let us consider the simpler case where the rates only depend on the variation of energy~(\ref{del}).

For the linear chain, the general form~(\ref{glau}) automatically verifies this requirement.
The so-called Glauber rate is obtained by fixing the parameter $\delta=0$.
It is the only such rate yielding linear equations for the temporal evolution of the observables.
For instance, the Metropolis rate 
\beqa\label{metro}
w(\s_n;\{\s_{j}\})=\min(1,\e^{-\Delta E/T})\nonumber\\
=\frac{2+\gamma}{2(1+\gamma)}\left(1-\frac{\gamma}{2+\gamma}(\s_{n-1}\s_{n+1}+\s_n(\s_{n-1}+\s_{n+1}))\right)
\eeqa
does not share this property.

For the square lattice, the requirement that the rates only depend on the variation of energy implies that the four neighbours of the central spin are equivalent, yielding 11 additional constraints
\beqa\label{onze}
c_4=c_3=c_2=c_1,\quad c_9=c_8=c_7=c_6, \nonumber\\
c_{15}=c_{14}=c_{13}=c_{12}=c_{11}=c_{10}.
\eeqa
The remaining independent coefficients are $c_0, c_1,c_5, c_6, c_{10}$.
The system of 16 equations $\{E_i=0\}$ for these 5 coefficients only gives two constraints
\beqa\label{delta}
c_1+\frac{\gamma}{2}(c_5-c_0)-c_6=0,
\nonumber\\
\frac{\gamma^2}{6}c_0-2\frac{1+\gamma^2}{3\gamma}c_1
-\frac{2+\gamma^2}{6}c_5-c_{10}=0.
\eeqa

From this general solution one can extract some simpler expressions for the rate function.
For instance, imposing $c_5=c_{10}=0$ yields
\beqa\label{glau2D}
w(\s_n;\{\s_{j}\})=\frac{\alpha}{2}\left(1-\frac{\gamma(2+\gamma^2)}{4(1+\gamma^2)}\s_n(\s_{j_1}+\s_{j_2}+\s_{j_{\underline 1}}+\s_{j_{\underline 2}})\right.
\nonumber\\
\left.
+\frac{\gamma^3}{4(1+\gamma^2)}\s_n(\s_{j_1}\s_{j_2}\s_{j_{\underline 1}}+\s_{j_2}\s_{j_{\underline 1}}\s_{j_{\underline 2}}+\s_{j_{\underline 1}}\s_{j_{\underline 2}}\s_{j_1}+\s_{j_{\underline 2}}\s_{j_1}\s_{j_2})\right)
\eeqa
which is the Glauber rate, usually written as
\beq\label{glau2D+}
w(\s_n;\{\s_{j}\})=\frac{\alpha}{2}\left(1-\s_n\tanh K(\s_{j_1}+\s_{j_2}+\s_{j_{\underline 1}}+\s_{j_{\underline 2}})\right).
\eeq

Another simple form is obtained by setting $c_1=c_5=0$:
\beqa
w(\s_n;\{\s_{j}\})=\frac{\alpha}{2}\left(1-\frac{\gamma}{2}\s_n(\s_{j_1}+\s_{j_2}+\s_{j_{\underline 1}}+\s_{j_{\underline 2}})\right.
\nonumber\\
\left.
+\frac{\gamma^2}{6}(\s_{j_1}\s_{j_2}+\s_{j_2}\s_{j_{\underline 1}}+\s_{j_{\underline 1}}\s_{j_{\underline 2}}+\s_{j_{\underline 2}}\s_{j_1}+\s_{j_2}\s_{j_{\underline 2}}+\s_{j_1}\s_{j_{\underline 1}})\right).
\eeqa

Let us finally note that the general form of a rate function satisfying detailed balance can be written as
\beq\label{qexpdet}
w(\s_n;\{\s_j\})=Q(\{\s_j\})\e^{-K\s_n h_n},
\eeq
where $Q(\{\s_j\})$ is a linear combination with arbitrary coefficients of the operators $O_i$ not containing the central spin.
This can be seen by multiplying both sides of~(\ref{db}) by $\e^{K\s_n h_n}$ and observing that the product 
$w(\s_n;\{\s_j\})\e^{K\s_n h_n}$ is even in $\s_n$.

For instance, for the linear chain,
\beq
Q(\{\s_j\})=a_0+a_3\,O_3,
\eeq
where $a_0$ and $a_3$ are arbitrary, and simply related to $c_0$ and $c_1$ of~(\ref{lin}): 
$a_0=(c_0-c_3+(c_0+c_3)/\cosh 2K)/2$, $a_3=(c_3-c_0+(c_0+c_3)/\cosh 2K)/2$.

For the square lattice, 
\beqa
Q(\{\s_j\})&=&a_0+a_5 O_5+\sum_{i=10}^{15}a_{i}\, O_{i}
\label{eq:Q}
\eeqa
depends on 8 arbitrary parameters.
If furthermore we ask that the rate function only depends on the variation of energy, then one should take
\beq\label{eq:detsym}
Q(\{\s_j\})=a_0+a_5\, O_5+a_{10}\sum_{i=10}^{15}O_i,
\eeq
with arbitrary coefficients $a_0,a_5,a_{10}$.
These are linearly related to the three independent parameters amongst $c_0,c_1,c_5, c_6, c_{10}$ related by~(\ref{delta}).
\section{Global balance}
\label{sec:global}

We now want to satisfy~(\ref{mast}), not term by term but as a whole.
As above we can solve the problem in either of two equivalent ways: by first finding the constraints on the rates $\{w_\alpha\}$, from which those on the coefficients $\{c_i\}$ ensue; or by first finding the constraints on the coefficients, from which those on the rates ensue.
Both ways have to be implemented by formal computations,
which are of equal algorithmic difficulty.
We start with the second one, of easier presentation.

\begin{table}[ht]
\caption{Number of constraints (or rank) for dynamics on regular lattices. First column: linear chain ($z=2$), square lattice ($z=4$), triangular lattice ($z=6$), cubic lattice ($z=6$).
The last line is for the hexagonal lattice (see text).
Second and third columns: number of equations and rank of the system of equations for detailed balance (db). Fourth and fifth columns: same for global balance (gb). 
Last column: number of free parameters for global balance.}
\label{tab:result}
\begin{center}
\begin{tabular}{|l||c|c||c|c|c|}
\hline
Lattice&$2^z$&Rank (db)&$M$&Rank (gb)&Free parameters (gb)\\
\hline
linear&$4$&$2$&$3$&$1$&$3$\\
square&$16$&$8$&$12$&$6$&$10$\\
triangular&$64$&$32$&$49$&$29$&$35$\\
cubic&$64$&$32$&$55$&$32$&$32$\\
\hline
hexagonal&8+8&8&12&8&$8$\\
\hline
\end{tabular}
\end{center}
\end{table}
\begin{table}[ht]
\caption{Number of constraints (or rank) and number of free parameters (gb) for dynamics on regular lattices at infinite temperature. 
Same examples as in Table~\ref{tab:result}.}
\label{tab:resultinfty}
\begin{center}
\begin{tabular}{|l||c||c|c|}
\hline
Lattice&Rank (db)&Rank (gb)&Free parameters (gb)\\
\hline
linear&$2$&$1$&$3$\\
square&$8$&$6$&$10$\\
triangular&$32$&$26$&$38$\\
cubic&$32$&$29$&$35$\\
\hline
hexagonal&$8$&$5$&$11$\\
\hline
\end{tabular}
\end{center}
\end{table}

\subsection{Constraints on the coefficients}
\label{sec:conscoef}

Remind that
\beq
B(\s_n;\{\s_j\})=\sum_{i=0}^{2^{z}-1}E_i\,O_i,
\eeq
where the $E_i$ are linear combinations of the $c_i$.
The detailed balance condition is just $\{E_i=0\}$ (see section~\ref{detbal}).
The sum in~(\ref{mast}) can be rewritten as\beq
\sum_{n}B(\s_n;\{\s_j\})=N \sum_i E_i\,\overline{O_i},
\eeq
defining the spatial averages of the spin operators as
\beq
\overline{O_i}=\frac{1}{N}\sum_{n}O_i.
\eeq
Taking into account the identities between the $\overline{O_i}$ due to translation invariance (see the examples below), the balance equation~(\ref{mast}) finally reads
\beq\label{eq:ej}
 \sum_{j} \widetilde{E_j}\,\overline{O_j}=0,
\eeq
where the $\overline{O_j}$ are a subset of the $\overline{O_i}$, and with $\widetilde{E_0}\equiv{E_0}$.
The size of this subset, i.e., the number $M$ of terms in this sum, is equal to the difference between $2^{z}$ and the number of identities due to translation invariance.
The $M$ equations $\{\widetilde{E_j}=0\}$ on the $c_i$ are no all independent a priori.
The rank of this system of equations is given in Table~\ref{tab:result} for the various examples that we now present.

\subsection{Examples}
\label{sec:exampl}

For the linear chain, we have, with the notation~(\ref{eq:oper1D}),
\beq
\overline{O_1}=\overline{O_2},
\eeq
hence~(\ref{eq:ej}) reads
\beq
{E_0}+({E_1}+E_2)\,\overline{O_1}+{E_3}\,\overline{O_3}=0.
\eeq
The $M=3$ equations $\{\widetilde{E_j}=0\}$ yield only one condition: $E_0=0$ (see~\ref{app:1D}).
In other words, the rank of this system of 3 equations is equal to 1 (see Table~\ref{tab:result}).
We find:
\beq\label{constraint1D1}
c_1+c_2+\gamma(c_0+c_3)=0,
\eeq
which generalizes the result~(\ref{lin}) found in the detailed balance case.
Hence, setting $c_2/c_0=\epsilon$, $c_3/c_0=\delta$ and $c_0=\alpha/2$,
the most general rate function satisfying the condition of global balance reads
\beqa\label{gener1}
w(\s_n;\{\s_j\})&=&
\frac{\alpha}{2}(1-(\gamma(1+\delta)+\epsilon)\,\s_{n}\s_{n+1}+\epsilon\,\s_{n-1}\s_{n}
\nonumber\\
&+&\delta\,\s_{n-1}\s_{n+1}),
\eeqa
which depends on the 3 arbitrary parameters $\alpha,\epsilon,\delta$.
The corresponding dynamics is asymmetric and irreversible.
Setting $\epsilon=-\gamma(1+\delta)(1-p)$, we can alternatively write:
\beqa\label{gener2}
w(\s_n;\{\s_j\})&=&
\frac{\alpha}{2}(1-\gamma(1+\delta)\s_{n}(p\s_{n-1}+(1-p)\s_{n+1})
\nonumber\\
&+&\delta\,\s_{n-1}\s_{n+1}).
\eeqa
The general Glauber form~(\ref{glau}) is recovered by setting $p=1/2$.
At the other end, the case of totally asymmetric dynamics where the central spin is only influenced by one of its neighbours leads, once a choice of neighbour is done, to a {\it unique} expression up to the time scale fixed by the coefficient $\alpha$.
For instance, if $\s_n$ is only influenced by its left neighbour, setting $\delta=0$ and $p=1$ ($\epsilon=-\gamma$), we obtain
\beq\label{rate:direct}
w(\s_n;\{\s_j\})=
\frac{\alpha}{2}\left(1-\gamma\,\s_{n-1}\s_{n}\right).
\eeq
Fixing the scale of time by the choice $\alpha=2\cosh 2 K$, we obtain the exponential form
\beq\label{kun1D}
w(\s_n;\{\s_j\})=\e^{-2 K\s_{n-1}\s_{n}}.
\eeq
We shall comment further, in section~\ref{positiv}, on the range of allowed parameters in~(\ref{gener1}) or~(\ref{gener2}).

\smallskip
\noindent{\it Remark }
Eq.~(\ref{constraint1D1}) can also be interpreted as the equation fixing the temperature of the model.
Hence, for the linear chain, any generic rate depending on the 4 parameters $c_0,c_1,c_2,c_3$ leads to a Gibbsian stationary measure.

\medskip
On the square lattice we have 4 identities due to translation invariance, which only involves two-spin operators,
\beq\label{2Dsym}
\overline{O_6}=\overline{O_9}, \quad\overline{O_7}=\overline{O_8}, 
\quad\overline{O_{10}}=\overline{O_{13}}, \quad\overline{O_{11}}=\overline{O_{14}}, 
\eeq
with the notations of Table~\ref{tab:op2D}.
The resulting system of $M=12$ $(16-4)$ linear equations $\{\widetilde{E_j}=0\}$, has rank 6, yielding 6 equations of constraint on the $c_i$.
In other words, 6 coefficients are expressed as linear combinations of the other 10 coefficients, which remain arbitrary.
Totally asymmetric cases are obtained by asking the rate function to depend only on two or three of the neighbouring spins instead of four.
For example keeping the east ($\s_{j_1}$) and north ($\s_{j_2}$) spins only, and cancelling the coefficients of the operators containing the two other spins $\s_{j_{\underline 1}}$ and $\s_{j_{\underline 2}}$, we obtain a {\it unique} rate function, up to a scale of time,
\beq\label{rate:ne}
w(\s_n;\{\s_j\})=\frac{\alpha}{2}
\left(1-\gamma\s_n(\s_{j_1}+\s_{j_2})+\gamma^2\,\s_{j_1}\s_{j_2}\right).
\eeq
Fixing this timescale by the choice $\alpha=2\cosh^2 2 K$, allows to write~(\ref{rate:ne}) into the exponential form\footnote{This form, as well as~(\ref{kun1D}), appear in~\cite{kun} without the factor 2. 
The corresponding stationary states have their temperature halved.}
\beq\label{kun2D}
w(\s_n;\{\s_j\})=
\e^{-2 K\s(\s_{j_1}+\s_{j_2})}.
\eeq

For the 3D cubic lattice, amongst the $2^z=64$ operators ${O_i}$, 18 are related, two by two, by translation invariance.
They all belong to the group of ${7\choose 2}=21$ two-spin operators.
The three two-spin operators of this group not related by translation invariance are
\beq
\s_{j_a}\s_{j_{\underline{a}}},
\eeq
for $a=1,2,3$, with notations analogous to those of Table~\ref{tab:op2D}.
So we have $M=55$ ($64-9$) linear equations $\{\widetilde{E_j}=0\}$ in the $\{c_i\}$ to solve.
The rank of this system of $M$ equations is equal to 32, as for the case of detailed balance.
The constraints are indeed found to be the same as when detailed balance holds (see also section~\ref{sub:constr} below).
On the cubic lattice there is no irreversible Gibbsian dynamics.

It is striking to compare the former case of the 3D cubic lattice to the case of the 2D triangular lattice, for which the coordination is the same ($z=6$).
For the triangular lattice, there are 15 identities due to translational invariance satisfied by the spatial averages of a subset of the $64$ operators ${O_i}$.
These 15 identities correspond to the translations of $\s_n\s_{j_a}$, for $a=1,2,3$ (9 relations), to the translations of $\s_{j_1}\s_{j_3}, \s_{j_2}\s_{j_{\underline{1}}},\s_{j_3}\s_{j_{\underline{2}}}$ (3 relations), and to the translations of the four-spin operators, e.g., $\s_n\s_{j_3}\s_{j_{\underline{1}}}\s_{j_{\underline{2}}}$ (3 relations).
Thus $M=49$ ($64-15$), and the rank of this system of equations is found to be equal to 29.
In this case, there do exist irreversible Gibbsian dynamics.

The totally asymmetric case involving the three spins $\s_{j_1}$, $\s_{j_2}$, $\s_{j_3}$ in the three unit directions is, again, determined {\it uniquely}, up to a time scale as
\beqa\label{ta:tri}
w(\s_n;\{\s_j\})=\frac{\alpha}{2}
\left(1-\gamma\s_n(\s_{j_1}+\s_{j_2}+\s_{j_3})\right.\nonumber\\
\left.+\gamma^2(\s_{j_1}\s_{j_2}+\s_{j_2}\s_{j_3}+\s_{j_1}\s_{j_3})
-\gamma^3\s_n\s_{j_1}\s_{j_2}\s_{j_3}
\right).
\eeqa
This rate function can also be written in exponential form as
\beq\label{kun2Dtri}
w(\s_n;\{\s_j\})=\e^{-2K \s_n(\s_{j_1}+\s_{j_2}+\s_{j_3})},
\eeq
with the choice $\alpha=2\cosh^3 2K$.

\subsection{Constraints on the rates}
\label{sub:constr}
One can, of course, deduce the constraints on the rates from the above.
Alternatively we can obtain these constraints directly by following the exact parallel of section~\ref{sec:conscoef}.
Remind that
\beq
B(\s_n;\{\s_j\})=\sum_{i=0}^{2^{z}-1} F_i\, O_i,
\eeq
where the $F_i$ are linear combinations of the $B_\alpha$.
The detailed balance condition is just $\{F_i=0\}$ (see section~\ref{detbal}).
The sum in~(\ref{mast}) can be rewritten as
\beq\label{eq:NFi}
\sum_{n}B(\s_n;\{\s_j\})=N \sum_i F_i\,\overline{O_i}.
\eeq
Taking into account the identities between the $\overline{O_i}$ due to translation invariance, the balance equation~(\ref{mast}) finally reads
\beq\label{eq:fj}
 \sum_{j} \widetilde{F_j}\,\overline{O_j}=0,
\eeq
where the $\overline{O_j}$ are a subset of the $\overline{O_i}$, and with $\widetilde{F_0}\equiv{F_0}$.
The $M$ equations $\{\widetilde{F_j}=0\}$ on the $B_\alpha$ are equivalent to the equations $\{\widetilde{E_j}=0\}$ on the $c_i$.
However they yield constraints on the rates, instead of constraints on the coefficients.

\subsection*{Remark}
One can recover~(\ref{eq:NFi}) following a slightly different path, as follows\footnote{This variant of the method was first introduced in~\cite{lg2006}.
}. 
For a given fixed configuration $\c$, the sum in~(\ref{mast}) can be rewritten as
\beq
\sum_{\alpha=1}^{2^z}\left(N_{\alpha}B_\alpha+\bar N_{\alpha}\bar B_\alpha\right),
\eeq
where $N_{\alpha}$ (resp. $\bar N_{\alpha}$) is the number of occurrences in $\c$ of the local configuration $\{$central spin up (resp. down) with the $z$ neighbours in configuration $\alpha$$\}$,
\beq
N_\alpha=\sum_n I(\s_n=+1;\{\s_j\}_\alpha),\quad \bar N_\alpha=\sum_n I(\s_n=-1;\{\s_j\}_\alpha).
\eeq
Using the spin symmetry and regrouping terms, the sum can be rewritten as
\beq\label{sum}
\sum_{\alpha=1}^{2^z}(N_{\alpha}+\bar N_{2^z+1-\alpha})B_\alpha.
\eeq
The numbers $N_\alpha$ can be decomposed on the basis of spatial averages of all the spin operators.
However, only the even operators remain in the sum $N_{\alpha}+\bar N_{2^z+1-\alpha}$ (see~(\ref{eq:decomp})).
Finally the sum~(\ref{sum}) yields~(\ref{eq:NFi}).

\subsection{Examples}

We illustrate the method on the cases considered above, in one to three dimensions.

For the linear chain, (\ref{eq:fj}) reads
\beq
{F_0}+({F_1}+F_2)\,\overline{O_1}+{F_3}\,\overline{O_3}=0.
\eeq
The 3 equations $\{\widetilde{F_j}=0\}$ yield only one condition: $F_0=0$, hence the detailed balance condition $B_1=w_1-\e^{-4K}\w_1=0$, i.e.,
\beq\label{constraint1D}
w(+;++)=\e^{-4K}w(-;++),
\eeq
which is equivalent to~(\ref{constraint1D1}).
The rank of this system of 3 equations is equal to 1, as already found above (see~\ref{app:1D}).

\smallskip
On the square lattice, taking into account the identities~(\ref{2Dsym}) due to translation invariance, the resulting system of $M=12$ linear equations in the $B_\alpha$, $\{\widetilde{F_j}=0\}$, has rank 6, yielding
the following 6 equations of constraint:
\beqa\label{six}
w_{1}-\e^{-8 K}\, \w_{1}=0,\nonumber \\
w_{6}-\w_{6}=0,\nonumber \\
w_{2}-\e^{-4 K}\, \w_{2}+w_{5}-\e^{-4 K}\, \w_{5}=0,\nonumber \\
\e^{4 K}\, w_{3}-\w_{3}-(w_{8}-\e^{4 K}\, \w_{8})=0,\nonumber \\
w_2-\e^{-4 K}\,\w_2-(w_3-\e^{-4 K}\,\w_3)
+\frac{2}{1+\e^{4 K}}(w_{7}-\w_{7})=0,\nonumber\\
w_2-\e^{-4 K}\,\w_2+w_3-\e^{-4 K}\,\w_3-\frac{2}{1+\e^{4 K}}(w_{4}-\w_{4})=0.
\eeqa
The space of independent parameters has dimension 10, in agreement with what was found above.

\smallskip
For the 2D triangular lattice, one recovers the results presented above: the 15 identities satisfied by the spatial averages of a subset of the $2^z=64$ operators ${O_i}$ yield a system of
$M=49$ equations, the rank of which is equal to 29.

\smallskip
For the 3D cubic lattice, we have $M=55$ ($64-9$) linear equations in the $B_\alpha$, ($\alpha=1,\ldots, 64$), to solve.
The constraints found are the 32 detailed balance conditions $B_\alpha=0$, in agreement with what was found above.

\subsection{Special case of the hexagonal lattice}

\begin{figure}
\begin{center}
\includegraphics[angle=0,width=0.6\linewidth]{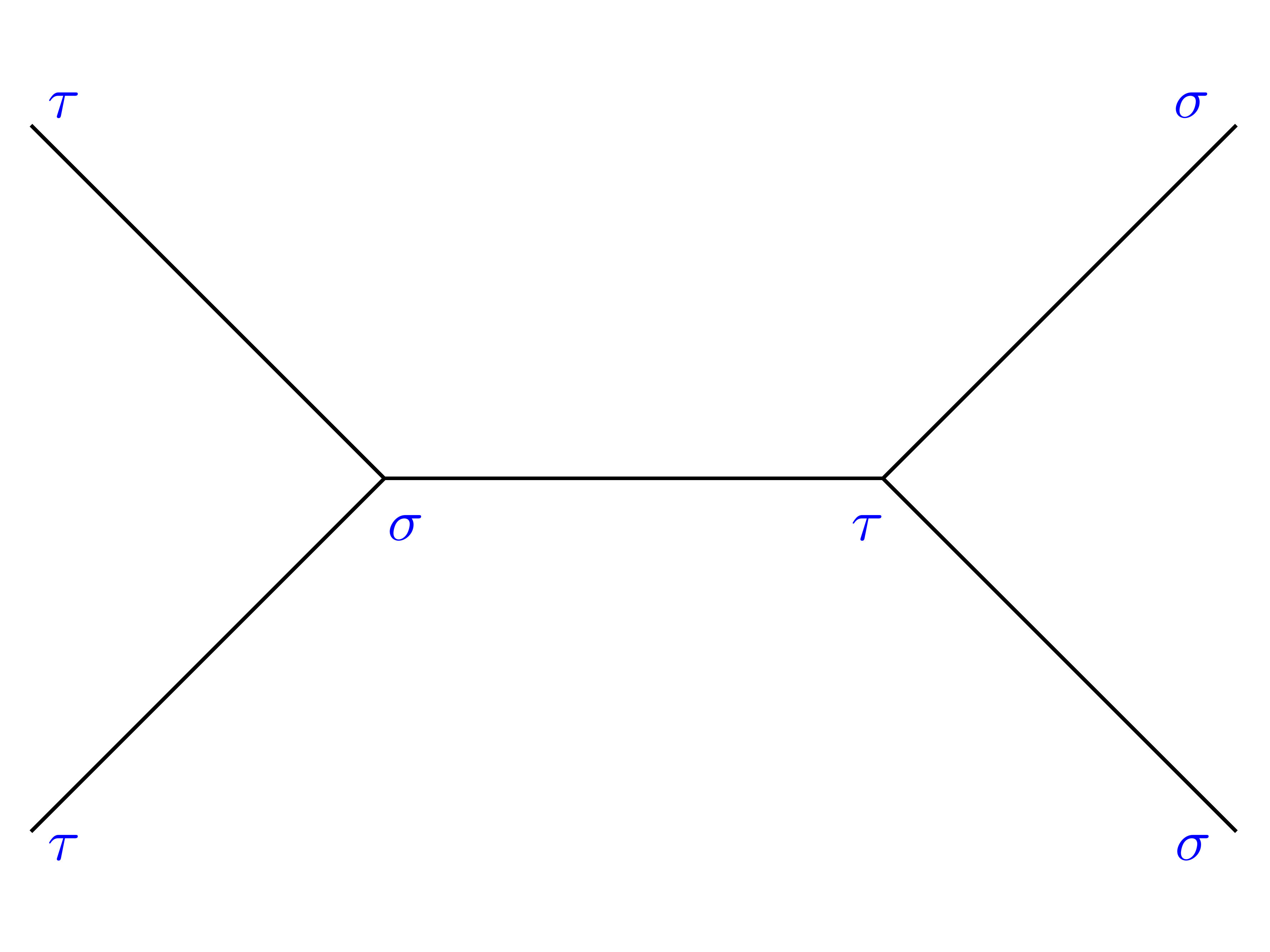}
\caption{\label{fig:hex}
Two types of environments for the hexagonal lattice.
}
\end{center}
\end{figure}

On the hexagonal lattice one has to distinguish two kinds of spins, named respectively $\s$ and $\tau$, corresponding to two types of environments (figure~\ref{fig:hex}).
Spin $\s_n$, located at $\x_n$, is surrounded by $\tau_{j_1}$, $\tau_{j_2}$ and $\tau_{j_3}$ with notations analogous to those of Table~\ref{tab:op2D},
where $\E_1$, $\E_2$ and $\E_3$ are the unit vectors spanning the hexagonal lattice.
Spin $\tau_n\equiv\tau_{j_1}$, located at $\y_n$, distant from $\x_n$ by one unit, is surrounded by $\s_{k_{\underline{1}}}$ (which is $\s_n$), 
$\s_{k_{\underline{2}}}$ and $\s_{k_{\underline{3}}}$ in the respective directions $-\E_1$, $-\E_2$ and $-\E_3$.
To these two spins correspond two rates
$w^{\s}(\s;\{\tau\})$ and $w^{\tau}(\tau;\{\s\})$, skipping the indices.
The balance equation~(\ref{eq:ej}) now reads
\beq\label{eq:ejhex}
\widetilde{E_0^{\s}}+\widetilde{E^{\tau}_0}+ \sum_{j} \left(\widetilde{E_j^{\s}}\,\overline{O_j^{\s}}
+\widetilde{E_j^{\tau}}\,\overline{O_j^{\tau}}\right)=0,
\eeq
where the operators ${O^{\s}}$ are the three $\s_n\tau_{j_a}$, the three $\tau_{j_a}\tau_{j_b}$ and $\s_n\tau_{j_1}\tau_{j_2}\tau_{j_3}$.
The operators ${O^{\tau}}$ are defined analogously.
In~(\ref{eq:ejhex}) the identities due to translation invariance have been taken into account:
\beq
\overline{\s_n\tau_{j_2}}=\overline{\tau_n\s_{k_{\underline{2}}}},\quad
\overline{\s_n\tau_{j_3}}=\overline{\tau_n\s_{k_{\underline{3}}}}.
\eeq
Moreover $\overline{\s_n\tau_{j_1}}$ and $\overline{\tau_n\s_{k_{\underline{1}}}}$ represent the same operator and $\widetilde{E_0^{\s}}+\widetilde{E^{\tau}_0}=0$ counts for one equation only.
The system of 12 resulting equations yield 8 constraints, identical to the detailed balance constraints.
Thus, on the hexagonal lattice, there is no irreversible Gibbsian dynamics, i.e., global balance enforces detailed balance, as long as temperature is finite (see below).

\subsection{Infinite temperature}

When temperature is infinite, the balance term takes the simpler form
\beq
B(\s_n;\{\s_{j}\})= w(\s_n;\{\s_{j}\})-w(-\s_n;\{\s_{j}\}),
\eeq
which only involves the operators containing the central spin.
Likewise, the only identities between operators due to translation invariance to be considered are those involving the central spin.

For instance, for the linear chain, $B(\s_n;\{\s_{j}\})$ only involves the two operators $O_1$ and $O_2$.
The identity $\overline{O_1}=\overline{O_2}$ still holds, thus the balance equation 
$(c_1+c_2)\overline{O_1}=0$ yields the constraint $c_1+c_2=0$, which is the limit of~(\ref{constraint1D1}) for $\gamma=0$.
The rate function reads
\beq
w(\s_n;\{\s_{j}\})=c_0+c_3O_3+c_1(O_1-O_2).
\eeq
The sum of the first two terms represent the infinite-temperature reversible rate function $Q(\{\s_j\})$ (see~(\ref{qexpdet})).
If furthermore the dynamics is totally asymmetric,
then one should impose $c_1=c_3=0$, i.e., $w(\s_n;\{\s_{j}\})=c_0$, in agreement with the infinite-temperature limit of~(\ref{rate:direct}).
In this limit the dynamics is reversible.

For the square lattice, $B(\s_n;\{\s_{j}\})$ only involves the 8 operators $O_1$ to $O_4$, and $O_6$ to $O_9$.
The identities to be considered are (see~(\ref{2Dsym}))
\beq
\overline{O_6}=\overline{O_9}, \quad\overline{O_7}=\overline{O_8}.
\eeq
The balance equation thus imposes the 6 constraints
\beq
c_1=c_2=c_3=c_4=0,\quad c_6+c_9=0,\quad c_7+c_8=0.
\eeq
This number of constraints is the same as at finite temperature.
The rate function reads
\beqa\label{eq:2Dinfty}
w(\s_n;\{\s_{j}\})&=&c_0+c_5O_5+\sum_{i=10}^{15}c_iO_i
\nonumber\\
&+&c_6(O_6-O_9)+c_7(O_7-O_8).
\eeqa
As above, the first line can be identified with the infinite-temperature reversible rate function $Q(\{\s_j\})$ defined in~(\ref{qexpdet}).
The second line is a linear combination of the operators $\s_n(\s_{j_a}-\s_{j_{\underline{a}}})$ ($a=1,2$).
If the dynamics is totally asymmetric, for instance keeping only the east and north spins,
then one should impose the vanishing of the coefficients corresponding to operators containing the west or south spins, i.e., $c_5,c_6,c_7$ as well as $c_{11}$ to $c_{15}$.
Thus
\beq\label{eq:2Dinfty+}
w(\s_n;\{\s_{j}\})=c_0+c_{10}O_{10},
\eeq
which corresponds to a reversible dynamics, as can be seen by comparing to the first line of~(\ref{eq:2Dinfty}).
This should be contrasted with the infinite-temperature limit of~(\ref{rate:ne}) which yields 
$w(\s_n;\{\s_{j}\})=c_0$.
There is no continuity of the finite-temperature result in this situation, when $T\to\infty$.

For the triangular lattice, the balance term involves 32 operators containing the central spin.
Only 6 symmetry relations remain, thus finally there are 26 constraints to satisfy. 
In other words there are more arbitrary parameters in the definition of the rate function satisfying global balance at infinite temperature than at finite temperature.
As for the square lattice, the rate is equal to the sum of $Q(\{\s_j\})$ ($32$ free parameters), of a linear combination of operator differences
$\s_n(\s_{j_a}-\s_{j_{\underline{a}}})$ ($a=1,2,3$) (3 free parameters), and of a linear combination of operator differences of the type $\s_n(\s_1\s_2\s_{\underline{3}}-\s_3\s_{\underline{1}}\s_{\underline{2}})$ (3 free parameters).
If the dynamics is totally asymmetric, for instance keeping only the spins in the direction $\E_1,\E_2,\E_3$, and taking into account the identities due to translation invariance, the resulting rate depends on 4 arbitrary coefficients, namely $c_0$ and the 3 coefficients corresponding to the 3 operators $\s_{j_1}\s_{j_2},\s_{j_1}\s_{j_3},\s_{j_2}\s_{j_3}$.
Again this dynamics is reversible but is different from the limit obtained from the finite-temperature result~(\ref{ta:tri}) at $T\to\infty$, which yields $w(\s_n;\{\s_{j}\})=c_0$.

For the cubic lattice, again the balance term involves 32 operators containing the central spin.
Only 3 symmetry relations remain, thus there are finally 29 constraints to satisfy.
Detailed balance is no longer enforced by global balance at infinite temperature.
The rate has the same form as for the triangular lattice, except that the four-spin operators do not enter the expression.
The case of totally asymmetric dynamics is identical to that found for the triangular lattice.

For the hexagonal lattice, as for the case of the cubic lattice, detailed balance is no longer enforced by global balance at infinite temperature.

Table~\ref{tab:resultinfty} summarizes the results for the various examples that we considered.

\section{Special forms of the rate function}

\subsection{Totally asymmetric dynamics}
For the totally asymmetric dynamics satisfying global balance encountered so far, the rate function could always be written in exponential form as
\beq\label{formexp}
w(\s_n;\{\s_j\})=\e^{-2K\s_n h^{+}_n},
\eeq
where $h_n=h_n^{+}+h_n^{-}$ is decomposed into two components related by inversion through the central spin
see~(\ref{kun1D}), (\ref{kun2D}) and (\ref{kun2Dtri}).
For instance on the square lattice, $h_n^{+}=\s_{j_1}+\s_{j_2}$, $h_n^{-}=\s_{j_{\underline 1}}+\s_{j_{\underline 2}}$.

Reciprocally, assume that the rate function has the form~(\ref{formexp})~\cite{gb2009}.
Then the balance term reads
\beq
B(\s_n;\{\s_j\})=\e^{-2K\s_n h^{+}_n}-\e^{-2K\s_n h^{-}_n}.
\eeq
Expanding the two exponential terms in the right side, the analysis confirms the fact that on 1D and 2D lattices such rates satisfy the balance equation~(\ref{mast}), but shows that this is not the case for the 3D cubic lattice, or more generally for lattices of coordination $z\ge8$~\cite{gb2009}. 

\subsection{A restricted representation of the rate function}

In preparation of the discussion of section~\ref{sec:oliv}, 
we now want to investigate whether there are representations of the rate function that generalize both the form~(\ref{formexp}) and the form encountered when detailed balance holds, namely~(\ref{qexpdet}).
Let us consider, along the lines of Ref.~\cite{oliv}, the following a priori representation of the rate function, instead of the general expression~(\ref{basis}):
\beq\label{woliv}
w(\s_n;\{\s_j\})=Q(\{\s_j\})\e^{-\s_n H_n},
\eeq
where $Q(\{\s_j\})$ is a linear combination of the operators $O_i$ not involving the central spin,
and where
\beq
H_n=\sum_{a=1}^{D}(A_a\s_{j_a}+A_{\underline{a}}\s_{j_{\underline{a}}}),
\eeq
and $A_a+A_{\underline{a}}=2K$.
We remind that $\s_{j_a}$ is the spin located at $\x_n+\E_a$, and $\s_{j_{\underline{a}}}$ is the spin located at $\x_n-\E_a$.
Setting
\beq
A_a=K+L_a,\quad A_{\underline{a}}=K-L_a,
\eeq
we can rewrite the rate function as
\beq
w(\s_n;\{\s_j\})=Q(\{\s_j\})\e^{-K\s_n h_n}\e^{-\s_n\sum_a L_a(\s_{j_a}-\s_{j_{\underline{a}}})}.
\eeq
If the $L_a$ vanish, one recovers the detailed balance form~(\ref{qexpdet}), made of the two first factors in the right side of this equation.
At the other end, if $L_a=\pm K$, and if $Q(\{\s_j\})$ is reduced to a constant, then one recovers~(\ref{formexp}).
However thus far this form is only an a priori representation of the rate function, on which 
one should now impose the global balance condition.
In so doing, we find the following:

\begin{description}
\item[1D] The form~(\ref{woliv}) is faithful and equivalent to the previous result~(\ref{gener1}).
The constraint of global balance is automatically encoded in this form, where
the prefactor $Q(\{\s_j\})=a_0+a_3\,O_3$, with arbitrary coefficients $a_0$ and $a_3$.

\item[2D] Let us take the example of the square lattice.
The constraints of global balance fixes 4 linear relations on the $\{a_j\}$ defining $Q(\{\s_j\})$ (see~(\ref{eq:Q})).
Thus the form~(\ref{woliv}) provides examples of rate functions satisfying global balance~\cite{oliv}.
The reciprocal is not true, i.e., any rate function satisfying global balance is not of the form~(\ref{woliv}).
For instance~(\ref{woliv}) implies the following relation amongst the rates:
\beq
\frac{\bar w_2}{w_2}=\e^{8K}\frac{w_5}{\bar w_5},
\eeq
which does not hold in general.

One could also have argued differently by noting that the form~(\ref{woliv}), which depends on 10 arbitrary parameters a priori, actually depends on 6 parameters once the 4 constraints on the $\{a_j\}$ are imposed, namely the four remaining arbitrary $\{a_j\}$ and the two parameters $L_1$ and $L_{2}$, while the general rate function obeying global balance depends on 10 arbitrary parameters (see section~\ref{sec:exampl}).
This form therefore only represents a subset of the most general rate functions obeying global balance.

\item[3D]
Since we already stated that there is no Gibbsian irreversible dynamics for the cubic lattice, {\it the form~(\ref{woliv}) cannot represent a rate function satisfying global balance}.
Since the constraint of global balance enforces detailed balance, it must necessarily suppress the dependence of $w(\s_n;\{\s_j\})$ in the parameters $L_a$. 
In other words, $Q(\{\s_j\})$ must vanish for any configuration of the neighbours $\{\s_j\}$ such that $\s_{j_a}\ne\s_{j_{\underline{a}}}$ for $a=1,2,3$.
For the configurations such that the equality holds, i.e., $\s_{j_a}=\s_{j_{\underline{a}}}$ for all $a=1,2,3$, detailed balance automatically holds, as it should.
This is illustrated by an explicit example in section~\ref{sec:oliv}.

\end{description}

\subsection{Interpolating schemes}

For the linear chain, (\ref{gener2}) gives an interpolation between the totally asymmetric cases and the symmetric one.

In similar fashion there are forms of the rate function in 2D, which interpolate between the totally asymmetric expression~(\ref{rate:ne}) and one of the forms valid under detailed balance.
For instance, for the square lattice,
\beqa
w(\s_n;\{\s_j\})&=&
\frac{\alpha}{2}\left(
1-\gamma\, p\,\s_n(\s_{j_1}+\s_{j_2})-\gamma\,(1-p)\,\s_n(\s_{j_{\underline 1}}+\s_{j_{\underline 2}})\right.
\nonumber\\
&+&\left.\gamma^2\,p\,\s_{j_1}\s_{j_2}+\gamma^2\,(1-p)\,\s_{j_{\underline 1}}\s_{j_{\underline 2}}
\right).
\label{rateP}
\eeqa
The totally asymmetric forms are obtained for $p=1$ or $0$.
For $p=1/2$, the rate function satisfies detailed balance, without being of the form~(\ref{eq:detsym}).
This expression is convenient when investigating the physical consequences of irreversibility for the two-dimensional Ising model with asymmetric dynamics~\cite{gp}.

\section{Review of Ref.~\cite{oliv} and the question of the rates for the cubic lattice}
\label{sec:oliv}
Reference~\cite{oliv} is chiefly concerned with the computation of the entropy production rate for irreversible Ising models with Gibbsian stationary states.
The rate function considered in this reference is of the form~(\ref{woliv}),
which is well adapted to the computation of the entropy production rate.
As said above, this choice is not restrictive in 1D, but does not account for the most general rate function satisfying global balance for the square lattice.
The expression obtained in~\cite{oliv} for the entropy production rate of the linear chain generalizes a result of~\cite{maes};
the computation of the entropy production rate of the square lattice is done for the totally asymmetric case~(\ref{kun2D}).

We now turn to the treatment of the 3D case given in~\cite{oliv}.
Ref.~\cite{oliv} claims that {\it there exist} irreversible Ising models on the cubic lattice, with Gibbsian stationary measure with respect to the Hamiltonian~(\ref{hamilt}), contradicting the results found in Ref.~\cite{gb2009} and recalled in the present work.
This statement is actually untrue and relies on an incomplete analysis, as we now demonstrate. (See also~\cite{erratum}.)

The rate function chosen in~\cite{oliv} for the 3D case is of the form~(\ref{woliv}) with the particular choice $L_1=L_2=L_3=L$, and where $Q(\{\s_j\})$ is a linear combinations of operators, not involving the central spin and satisfying some additional symmetry requirements, with coefficients named $b_0,b_1,\ldots, b_5$ (see eq.~(67) in~\cite{oliv}).
It is an easy task to solve the problem of determining these unknown coefficients if one imposes the global balance condition, using the general methods described in section~\ref{sec:global}.
We find the following constraints on the coefficients $b_0,b_1,\ldots, b_5$,
\beq\label{resoliv}
b_5=b_0,\quad b_4=b_2=\frac{3 b_1-b_0}{2},\quad b_3=b_1,
\eeq 
independently of the value of the parameter $L$\footnote{We also checked that the equations of constraint on $b_0,b_1,\ldots, b_5$ written in~\cite{oliv} lead to the result~(\ref{resoliv}). In~\cite{oliv} the determination of these equations of constraint is done by identification of two forms of the balance term, the first one deduced from the choice made for the rate $w$, the other one is a linear combination of a subset of the operators $O_i$ chosen according to some symmetry requirements.
}.
One can then check that, taking into account the constraints~(\ref{resoliv}), the rate function thus obtained either satisfies the detailed balance condition, when $\s_{j_1}+\s_{j_2}+\s_{j_3}=\s_{j_{\underline{1}}}+\s_{j_{\underline{2}}}+\s_{j_{\underline{3}}}$, or vanishes when this condition does not hold, because the prefactor $Q(\{\s_j\})$ vanishes itself.
This is equivalent to saying that, for this choice of rate function, the condition of global balance enforces the condition of detailed balance.

These results are in agreement with the general statement, made in~\cite{gb2009} and in section~\ref{sec:global}, that there are no Gibbsian irreversible models for the cubic lattice, and therefore no entropy production for dynamics of the form~(\ref{woliv}) in this case.
\section{Positivity of the rates}
\label{positiv}
The rate functions found by the method above must satisfy the additional constraint of positivity for the various possible configurations.

\begin{figure}
\begin{center}
\includegraphics[angle=0,width=1\linewidth]{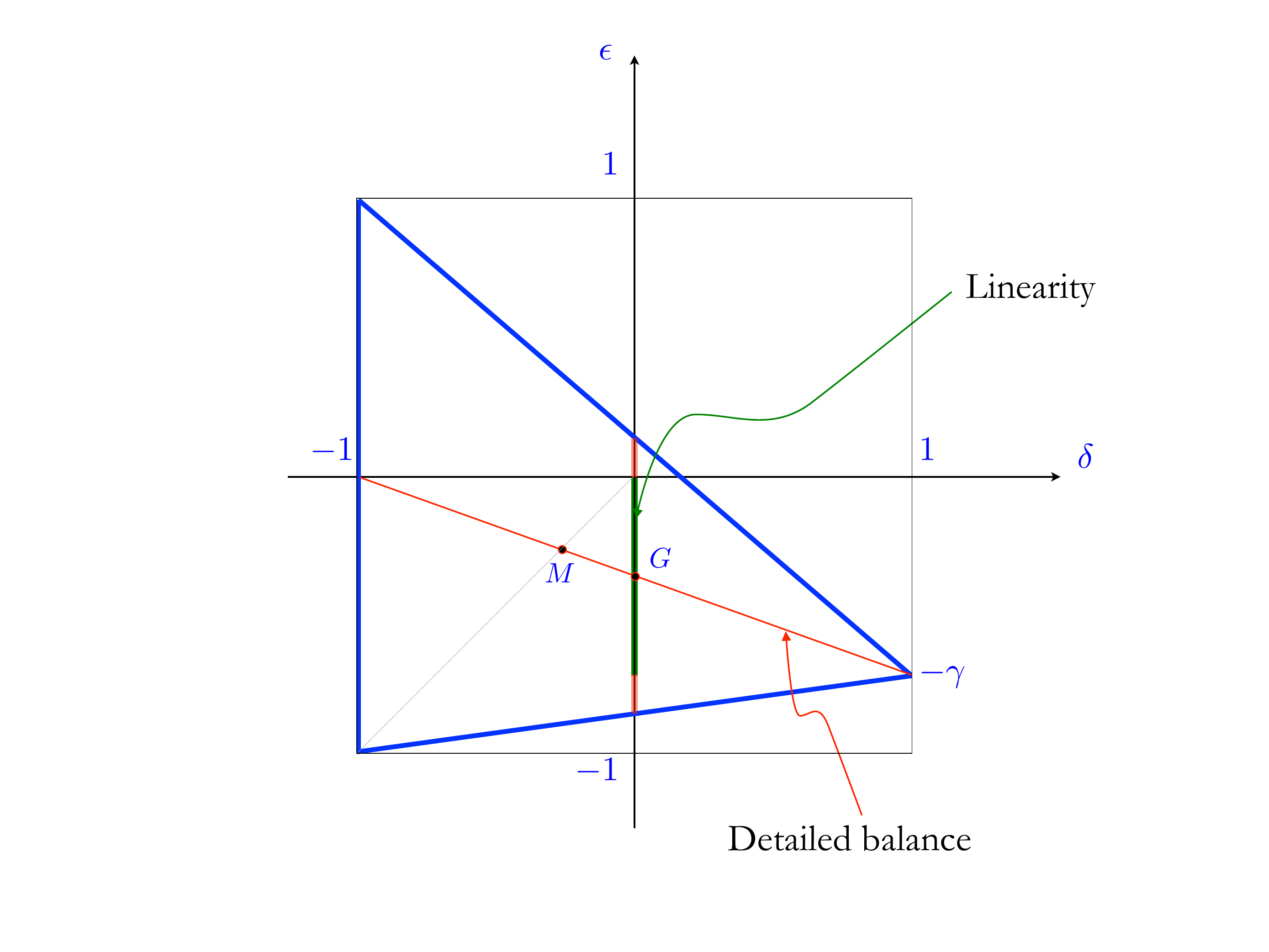}
\caption{\label{fig:visu}
Constraints due to positivity for the linear chain: allowed values of the parameters yielding positive rates and satisfying global balance are inside the triangle.
}
\end{center}
\end{figure}

We illustrate the issue on the 1D case, using~(\ref{gener1}).
The allowed region in the plane of the two parameters ($\delta$, $\epsilon$) yielding positive rates is the triangle depicted in figure~\ref{fig:visu}.
The sides of the triangle correspond to the vanishing of one of the rates $w_\alpha$.
The segment joining the two points $(-1,-1)$ and $(1,-\gamma)$ corresponds to $w_2\equiv w(+;+-)=0$.
The segment joining the two points $(-1,1)$ and $(1,-\gamma)$ corresponds to $w_3\equiv w(+;-+)=0$.
Finally the vertical segment at $\delta=-1$ corresponds to $w_1=w_4=0$, i.e., $w(+;++)=w(+;--)=0$.

All rates on the line joining $(-1,0)$ to $(1,-\gamma)$ satisfy detailed balance.
For example, the point marked $M$ corresponds to the Metropolis rate~(\ref{metro}).
All rates with $\delta=0$ lead to linear equations for the temporal evolution of the observables~\cite{cg2011}.
The point $G$, located at $(0,-\gamma/2)$ and corresponding to the Glauber rate, is the only point where both detailed balance and linearity hold. 

The two ends of the green segment are the totally asymmetric points $(0,-\gamma)$ and $(0,0)$, corresponding  respectively to values of the interpolating parameter $p=0$ and $p=1$ in~(\ref{gener2}).
The range of allowed values with $\delta=0$ goes beyond this segment.
It is comprised between 
the two extreme points $(0,-(1+\gamma)/2)$ and $(0,(1-\gamma)/2)$ which correspond respectively to the values $p=(\gamma-1)/(2\gamma)$, which is negative, and $p=(\gamma+1)/(2\gamma)$, which is larger than 1.
For those points and more generally for the range of values depicted in red in figure~\ref{fig:visu} the magnetization of the linear chain exhibits an oscillating relaxation~\cite{tocome}.
For instance, for the point $(0,-(1+\gamma)/2)$, the rate function~(\ref{gener1}) becomes
\beq
w(\s_n;\{\s_j\})=
\frac{\alpha}{2}\left(1+\s_n\Big(\frac{1-\gamma}{2}\s_{n+1}-\frac{1+\gamma}{2}\s_{n-1}\Big)\right).
\eeq

%

%

\section{Discussion}

Let us come back on the interplay between irreversibility and asymmetry of the dynamics in the present context.
We recall that the dynamics is symmetric if the rates are given by a symmetric function of the neighbouring spins $\{\s_j\}$ of the flipping spin $\s_n$.
In other words under symmetric dynamics the neigbhouring spins have equal influence on the flipping spin.
We ask:
\begin{enumerate}
\item Can a dynamics be both irreversible and symmetric? 

\item Can a dynamics be both reversible and asymmetric? 

\end{enumerate}

A negative answer to the first question means that irreversibility necessarily implies asymmetry.
A negative answer to the second question means that reversibility necessarily implies symmetry, which is the reciprocal of (i). We illustrate the issue on the examples of the linear chain and of the square lattice.

For the linear chain the answer to the two questions is negative.
Reversibility and symmetry are equivalent.
Symmetry of the dynamics for the linear chain requires $c_1=c_2$ (see~(\ref{lin})), which appears also as a constraint imposed by the condition of detailed balance.

For the square lattice, the answer to the two questions is positive.
\begin{enumerate}
\item Firstly, symmetry of the dynamics does not imply reversibility.
Indeed, starting from the generic rate function depending on 16 parameters, if one imposes symmetry, then only 5 independent parameters remains, which are the coefficients $c_0,c_1,c_5,c_6,c_{10}$ (see~(\ref{onze})).
If no further condition is imposed, the dynamics is generically irreversible.
The voter model~\cite{voter} and the broader class of models defined by the rate function~\cite{p1p20,p1p2}
\beq
w(\s_n;\{\s_j\})=\frac{\alpha}{2}(1-\s_n\tanh\left[ \beta(h_n)\,h_n\right]),
\eeq
where $h_n=\sum_j\s_j$ (choosing $J=1$), provide examples of such a situation.
The inverse temperature takes three values, according to the value of the local field $h_n$: $\beta(0)$, $\beta(2)=\beta(-2)$ and $\beta(4)=\beta(-4)$.
This rate function is clearly symmetric in the neighbouring spins.
It can be rewritten as
\beqa
w(\s_n;\{\s_j\})&=&\frac{\alpha}{2}\left(1+\frac{1}{4}\big(\gamma_2-\frac{\gamma_4}{1+\gamma_4^2}\big)
(O_1+O_2+O_3+O_4)\right.
\nonumber\\
&-&\left.\frac{1}{4}\big(\gamma_2+\frac{\gamma_4}{1+\gamma_4^2}\big)
(O_6+O_7+O_8+O_9)\right),
\eeqa
where $\gamma_{2}=\tanh 2\beta(2)$, $\gamma_{4}=\tanh 2\beta(4)$.
Note that we have $c_5=c_{10}=0$.
These models correspond to genuinely irreversible dynamics since the rates do not even satisfy the constraints of global balance~(\ref{six}),
hence their stationary measure is unknown.
The voter model corresponds to the choice $\gamma_2=1/2$, $\gamma_4=1$.
The noisy voter model corresponds to the choice $\gamma_2=\gamma_4/(1+\gamma_4^2)$.
The Glauber rate function~(\ref{glau2D}) or (\ref{glau2D+}) is recovered by fixing $\gamma_2=\gamma_4=\gamma$.

Let us finally mention that imposing global balance on a generic symmetric rate function yields the two constraints~(\ref{delta}), i.e., reversibility is recovered.
In other words, irreversible Gibbsian dynamics are necessarily asymmetric.

\item
Secondly, reversibility does not imply symmetry of the dynamics.
Indeed any generic reversible rate function depends on 8 parameters (see for example~(\ref{eq:Q})).
On the other hand the number of independent parameters corresponding to reversible symmetric dynamics is equal to 3 (see~(\ref{eq:detsym})).
Thus generically any rate function satisfying detailed balance is asymmetric.
For instance the rate function
\beqa
w(\s_n;\{\s_j\})&=&\frac{\alpha}{2}\left(1-\frac{\gamma}{2}\s_n(\s_{j_1}+\s_{j_2}+\s_{j_{\underline 1}}+\s_{j_{\underline 2}})\right.
\nonumber\\
&+&\left.\frac{\gamma^2}{2}(\s_{j_1}\s_{j_{\underline 1}}+\s_{j_2}\s_{j_{\underline 2}})\right)
\eeqa
provides an example of a reversible asymmetric process.
In particular $w_4$ is not equal to $w_6$, as would be the case for a fully symmetric dynamics.
Eq.~(\ref{rateP}) is another example where, when $p=1/2$, the dynamics is reversible but not fully symmetric.
Yet another example is 
\beqa
w(\s_n;\{\s_j\})&=&\frac{\alpha}{2}\left(1-\frac{\gamma}{2-\gamma^2}\s_n(\s_{j_1}+\s_{j_2}+\s_{j_{\underline 1}})\right.
\nonumber\\
&+&\frac{\gamma^2}{2-\gamma^2}(\s_{j_1}\s_{j_2}+\s_{j_1}\s_{j_{\underline 1}}+\s_{j_2}\s_{j_{\underline 1}})
\nonumber\\
&-&\left.\frac{\gamma^3}{2-\gamma^2}\s_n\s_{j_1}\s_{j_2}\s_{j_{\underline 1}}
-\frac{1-\gamma^2}{2-\gamma^2}\s_n\s_{j_{\underline 2}}
\right),
\eeqa
which illustrates the fact that, even for reversible dynamics, one of the neighbouring spins (here $\s_{j_{\underline 2}}$) can play a role different from the other ones.

It is however easy to convince oneself, using~(\ref{qexpdet}), that reversibility and total asymmetry are incompatible, except at infinite temperature, as demonstrated by~(\ref{eq:2Dinfty+}).

\end{enumerate}

%
%
\section{Conclusion}

The present work is a completion of~\cite{gb2009}.
One of the questions raised and solved in this reference concerned the possible existence of irreversible single-spin flip dynamics with Gibbsian stationary states for ferromagnetic Ising systems.
The motivation was twofold. 

On the one hand, a natural question raised by the examples given in the past by K\"{u}nsch~\cite{kun} for totally asymmetric dynamics in one and two dimensions, is to what extent are these examples unique, and can they be extended to higher dimensions than two.
The result of~\cite{gb2009}, completed here, is that, as long as the dynamics is not totally asymmetric, the space of parameters defining the rate function allowing irreversible Gibbsian Ising models is large (see Table~\ref{tab:result}).
However, imposing total asymmetry of the dynamics yields a unique solution, up to a time scale, for the examples considered (linear chain, square and triangular lattices).
The answer to the second part of the question is presumably negative. Indeed, firstly, there is no such Gibbsian irreversible dynamics for the cubic lattice; secondly, one can argue that there are neither totally asymmetric Gibbsian dynamics for lattices of coordination $z\ge8$~\cite{gb2009}.
A novel outcome of the present work is that the situation can be different at infinite temperature (see~Table~\ref{tab:resultinfty}).

On the other hand, the models thus defined are interesting laboratories for the study 
of the physical consequences of irreversibility, in particular of the properties of the resulting nonequilibrium stationary state.
For instance, for the linear chain, though the stationary measure is Boltzmann-Gibbs, the dynamical properties of the relaxing system are changed~\cite{cg2011}.
Irreversibility also implies a non-vanishing entropy production rate
in the stationary state which can be exactly computed for irreversible Gibbsian models since the stationary measure is known~\cite{oliv,maes}.

The method used in~\cite{gb2009} and in the present work for the solution of the question raised above relies on linear algebra and properties of the system under translations.
This implies solving the system of linear equations of constraint on the rates by a formal computation.
It would be desirable to answer the same question by other means which would in some sense generalize the argument recalled above for lattices with coordination number $z\ge 8$.

\ack It is a pleasure to thank J.M. Luck and S. Prolhac for
helpful discussions.
I wish to acknowledge M.J. de Oliveira for exchange of correspondence and for gracefully agreeing with the analysis of section 6~\cite{erratum}.

\appendix

\section{Linear chain}
\label{app:1D}

Hereafter we give some details on the case of the linear chain.

We use the identity
\beq
\e^{-2K\s_n h_n}=\frac{1}{1-\gamma^2}(1-\gamma O_1)(1-\gamma O_2)
\eeq
($\gamma=\tanh 2K$) to obtain the decomposition~(\ref{ee}) of the balance term $B(\s_n;\{\s_j\})$ on the basis of operators $O_i$.
The coefficients $E_i$ thus obtained are, up to a global constant equal to $1/(1-\gamma^2)$,
\beqa\label{app:E}
E_0=-\gamma(c_1+c_2+\gamma(c_0+c_3)),
\nonumber\\
E_1=2c_1-\gamma^2(c_1-c_2)+\gamma(c_0+c_3),
\nonumber\\
E_2=2c_2+\gamma^2(c_1-c_2)+\gamma(c_0+c_3),
\nonumber\\
E_3=E_0,
\eeqa
from which the relation $E_0/\gamma+(E_1+E_2)/2=0$ is seen to hold.
Hence the rank of the system $\{E_i=0\}$ is equal to 2.

The matrix $A=(a_{i,\alpha})$ with $0\le i\le3$, and $1\le \alpha\le4$, which relates the coefficients $c_i$ to the rates $w_\alpha$, reads
\beq
A=\frac{1}{4}\pmatrix{
1&1&1&1
\cr
1&-1&1&-1
\cr
1&1&-1&-1
\cr
1&-1&-1&1
}.
\eeq
Hence, using~(\ref{Fi}), we have, for the coefficients $F_i$ defined in~(\ref{eq:ff}),
\beqa
F_0=(B_1+B_2+B_3+B_4)/4,
\nonumber\\
F_1=(B_1-B_2+B_3-B_4)/4,
\nonumber\\
F_2=(B_1+B_2-B_3-B_4)/4,
\nonumber\\
F_3=(B_1-B_2-B_3+B_4)/4.
\eeqa
Noting that $B_2+B_3=0$, we obtain, up to a global constant equal to $1/4$,
\beqa\label{app:F}
F_0=(1-\e^{4K})(w_1-\e^{-4K}\bar w_1),
\nonumber\\
F_1=(1+\e^{4K})(w_1-\e^{-4K}\bar w_1)-2(w_2-\bar w_2),
\nonumber\\
F_2=(1+\e^{4K})(w_1-\e^{-4K}\bar w_1)+2(w_2-\bar w_2),
\nonumber\\
F_3=F_0,
\eeqa
thus we have the relation $F_0/\gamma+(F_1+F_2)/2=0$.
Hence the rank of the system $\{F_i=0\}$ is equal to 2.

The two writings~(\ref{app:E}) and (\ref{app:F}) can be identified by using~(\ref{oper1D}), i.e.,
\beq
w(\s_n;\{\s_{j}\})=c_0+c_1\,\s_{n}\s_{n+1}+c_2\,\s_{n-1}\s_{n}+c_3\,\s_{n-1}\s_{n+1}.
\eeq

%
%
\section{Asymmetric Gibbsian conserved dynamics for the linear chain}
In this appendix we use the methods of the present paper to determine the rate function when the dynamics is conserved and satisfies global balance. 
So doing we recover the results of a previous work, which were established by a variant of the present method, and written differently~\cite{lg2006}.

\subsection{Basic facts}
The dynamics of the chain consists in flipping a bond chosen at random, say bond $(\s_n,\s_{n+1})$, if the two spins $\s_n$ and $\s_{n+1}$ are anti-aligned:
either $+-$ flips into $-+$, or $-+$ flips into $+-$.
The change in energy is equal to
\beq
\Delta E=2J(\s_{n-1}\s_{n}+\s_{n+1}\s_{n+2}).
\eeq
This is done with a rate $w(\s_n,\s_{n+1};\{\s_j\})$, where $\{\s_j\}$ is a notation for the two neighbours $\s_{n-1}$ and $\s_{n+2}$.
The number of values taken by the rate function is therefore equal to 8.
We denote the $4$ rates with $(\s_n=+1,\s_{n+1}=-1)$ by $w_{\alpha}$ and the other $4$ rates, corresponding to $(\s_n=-1,\s_{n+1}=+1)$, by $\bar w_{\alpha}$:
\beq
\label{walc}
w_{\alpha}=w(+-;\{\s_j\}_\alpha),
\qquad \bar w_\alpha=w(-+;\{\s_j\}_\alpha),
\eeq
(see Table~\ref{tab:1dconserved}).

\begin{table}[ht]
\caption{List of local configurations and corresponding values of the rate function for the one-dimensional chain with conserved dynamics.
There are 4 possible rates $w_{\alpha}$, with $(\s_n=+1,\s_{n+1}=-1)$, corresponding to the 4 possible configurations $\{\s_j\}$, labelled by $\alpha$, of the two neighbours of the flipping bond, taken in the order: left, right.
The 4 remaining rates $\w_\alpha$ correspond to $(\s_n=-1,\s_{n+1}=+1)$.}
\label{tab:1dconserved}
\begin{center}
\begin{tabular}{|c||c|c||c|c|}
\hline
$\alpha$&$\s_n,\s_{n+1};\{\s_j\}$&$w_\alpha$&$\s_n,\s_{n+1};\{\s_j\}$&$\w_\alpha$\\
\hline
$1$&$+-;++$&$w_{1}$&$-+;++$&$\w_1$\\
$2$&$+-;{+-}$&$w_2$&$-+;+-$&$\w_2$\\
$3$&$+-;{-+}$&$w_3$&$-+;-+$&$\w_3$\\
$4$&$+-;{--}$&$w_4$&$-+;--$&$\w_4$\\
\hline
\end{tabular}
\end{center}
\end{table}
\begin{table}[ht]
\caption{List of operators made of the 4 spins $\s_{n-1},\ldots,\s_{n+2}$.}
\label{tab:op}
\begin{center}
\begin{tabular}{|l|l|}
\hline
$i$&$O_i$\\
\hline
$1$&$\s_{n-1}\s_{n}\s_{n+1}\s_{n+2}$\\
$2$&$\s_{n-1}\s_{n}\s_{n+1}$\\
$3$&$\s_{n-1}\s_{n}\s_{n+2}$\\
$4$&$\s_{n-1}\s_{n+1}\s_{n+2}$\\
$5$&$\s_{n}\s_{n+1}\s_{n+2}$\\
$6$&$\s_{n-1}\s_{n}$\\
$7$&$\s_{n-1}\s_{n+1}$\\
$8$&$\s_{n-1}\s_{n+2}$\\
$9$&$\s_{n}\s_{n+1}$\\
$10$&$\s_{n}\s_{n+2}$\\
$11$&$\s_{n+1}\s_{n+2}$\\
$12$&$\s_{n-1}$\\
$13$&$\s_{n}$\\
$14$&$\s_{n+1}$\\
$15$&$\s_{n+2}$\\
\hline
\end{tabular}
\end{center}
\end{table}

Let us introduce the basis of $16$ spin operators $O_1,\ldots,O_{15}$, made of the 4 spins $\s_{n-1},\ldots,\s_{n+2}$, with $O_0=1$ (see Table~\ref{tab:op}).
We define the indicator variables
\beqa
I_\alpha=I(+-;\{\s_j\}_\alpha)=\frac{1+\s_n}{2}\frac{1-\s_{n+1}}{2}J_\alpha,
\nonumber\\
\bar I_\alpha= I(-+;\{\s_j\}_\alpha)=\frac{1-\s_n}{2}\frac{1+\s_{n+1}}{2}J_\alpha,
\eeqa
where $J_\alpha$ denotes the indicator variable of the event $\{$$\{\s_j\}$ in configuration $\alpha$$\}$.
We thus have, using the notation~(\ref{walc}),
\beq
w(+-;\{\s_{j}\})=\sum_{\alpha=1}^{{4}}J_\alpha\,w_\alpha,
\quad
w(-+;\{\s_{j}\})=\sum_{\alpha=1}^{{4}}J_\alpha\,\bar w_\alpha.
\eeq
Alternatively, we have
\beqa\label{app:dec1}
w(+-;\{\s_{j}\})&=&c_0 +c_8 O_8+c_{12} O_{12}+c_{15} O_{15},
\nonumber\\
w(-+;\{\s_{j}\})&=&d_0 +d_8 O_8+d_{12} O_{12}+d_{15} O_{15}.
\eeqa
Finally we can write the rate function as
\beqa\label{app:dec2}
w(\s_n,\s_{n+1};\{\s_j\})&=&
\frac{1+\s_n}{2}\frac{1-\s_{n+1}}{2}w(+-;\{\s_{j}\})
\nonumber\\
&+&\frac{1-\s_n}{2}\frac{1+\s_{n+1}}{2}w(-+;\{\s_{j}\}).
\eeqa

The balance term reads
\beqa\label{app:bal}
B(\s_n,\s_{n+1};\{\s_j\})&=&w(\s_n,\s_{n+1};\{\s_j\})
\nonumber\\
&-&w(-\s_n,-\s_{n+1};\{\s_j\})\e^{-\Delta E/T},
\eeqa
with
\beq\label{app:dec3}
\e^{-\Delta E/T}=\frac{1}{1-\gamma^2}(1-\gamma\, O_6)(1-\gamma\, O_{11}),
\eeq
($\gamma=\tanh 2K$).
Using~(\ref{app:dec1}), (\ref{app:dec2}) and (\ref{app:dec3}), we obtain a decomposition of the balance term~(\ref{app:bal}) on the basis of operators of Table~\ref{tab:op}:
\beq
B(\s_n,\s_{n+1};\{\s_j\})=E_0+\sum_{i=1}^{15}E_i\,O_i.
\eeq

\subsection{Symmetries}

We analyze the constraints induced on the rate function by the two following symmetries.

\subsection*{Symmetry under $P$, the spatial left-right parity}
The constraint induced by this symmetry reads
\beq
w(+-;\s_{n-1},\s_{n+2})=w(-+;\s_{n+2},\s_{n-1}),
\eeq
which imposes 
\beq
d_0=c_0,\ d_8=c_8,\ d_{12}=c_{15},\ d_{15}=c_{12},
\eeq
or
\beq
w_1=\w_1,\quad w_2=\w_3,\quad w_3=\w_2,\quad w_4=\w_4.
\eeq

\subsection*{Symmetry under $CP$}
This symmetry is the product of $C$ and $P$, where the charge conjugation $C$ changes the spins into their opposites.
In other words, the rates are the same for $+$ going to the right or for $-$ going to the left, with the environment of the latter conjugated to the environment of the former.
The constraints induced by this symmetry read
\beqa
w(+-;\s_{n-1},\s_{n+2})=w(+-;-\s_{n+2},-\s_{n-1}),
\nonumber\\
w(-+;\s_{n-1},\s_{n+2})=w(-+;-\s_{n+2},-\s_{n-1}).
\eeqa
This fixes 
\beq\label{app:cp}
c_{12}+c_{15}=0,\quad d_{12}+d_{15}=0,
\eeq
or 
\beq\label{app:cp2}
w_1=w_4,\quad \w_1=\w_4.
\eeq

\subsection{Detailed balance}
\label{sec:app:db}
The detailed balance condition imposes $B(\s_n,\s_{n+1};\{\s_j\})=0$, i.e., $E_i=0$ for all $i$.
We thus obtain the 4 constraints
\beqa\label{app:db}
d_0=\frac{1}{1-\gamma^2}(c_0-\gamma(\gamma c_8-c_{12}+c_{15})),\nonumber\\
d_8=\frac{1}{1-\gamma^2}(c_8-\gamma(\gamma c_0+c_{12}-c_{15})),\nonumber\\
d_{12}=\frac{1}{1-\gamma^2}(c_{12}-\gamma(\gamma c_{15}-c_{0}+c_{8})),\nonumber\\
d_{15}=\frac{1}{1-\gamma^2}(c_{15}-\gamma(\gamma c_{12}+c_{0}-c_{8})),
\eeqa
which express the equalities
\beqa\label{app:dbw}
w_1=\w_1,\quad w_2=\w_2\,\e^{-4K},\quad w_3=\w_3\,\e^{4K},\quad w_4=\w_4.\quad 
\eeqa

We now restrict the rate function furthermore by symmetry requirements.
\subsection*{Symmetry under $P$, the spatial left-right parity}
The additional constraint induced by this symmetry is
\beq
\gamma (c_0-c_8)+c_{12}-c_{15}=0,
\eeq
which expresses the equality $w_2=\e^{-4K}w_3$.
The resulting rate function can be read off from~(\ref{app:dec1}) and~(\ref{app:dec2}):
\beqa\label{app:sym}
w(+-;\{\s_j\})&=&
c_0+c_8\s_{n-1}\s_{n+2}+c_{12}\s_{n-1}+(\gamma (c_0-c_8)+c_{12})\s_{n+2},
\nonumber\\
w(-+;\{\s_{j}\})&=&c_0+c_8\s_{n-1}\s_{n+2}+c_{12}\s_{n+2}+(\gamma (c_0-c_8)+c_{12})\s_{n-1}.
\nonumber\\
\eeqa
It depends on 3 arbitrary coefficients.

\subsection*{Symmetry under $CP$}
The constraints on the rate function are the three first lines of~(\ref{app:db}) with $c_{12}+c_{15}=0$ and
$d_{15}$ is fixed equal to $-d_{12}$.
Again the resulting rate function depends on 3 free coefficients.

\subsection{Global balance}
We now turn to the global balance condition.
Translation invariance imposes
\beqa
\overline{O_2}=\overline{O_5},\qquad \overline{O_6}=\overline{O_9}=\overline{O_{11}},\qquad \overline{O_7}=\overline{O_{10}},\nonumber\\
\overline{O_{12}}=\overline{O_{13}}=\overline{O_{14}}=\overline{O_{15}}.
\eeqa
Solving the system of equations $\widetilde{E_j}=0$ (see~(\ref{eq:ej})) yields 2 constraints:
\beqa\label{app:res}
d_8=\frac{1}{1-\gamma^2}(c_8-\gamma(\gamma c_0+c_{12}-c_{15})),\nonumber\\
\gamma d_0-d_{12}+d_{15}=\frac{1}{1-\gamma^2}(-c_{12}+c_{15}-\gamma (c_0-(2-\gamma^2)c_8)),
\eeqa
which express the relations between rates
\beqa\label{app:res2}
w_2-\w_2\e^{-4K}+\w_3-w_3\e^{-4K}=0
\nonumber\\
w_1-\w_1+w_4-\w_4+\w_2(1+\e^{-4K})-w_2(1+\e^{4K})=0.
\eeqa
The number of free coefficients is equal to 6.

\subsection*{Symmetry under $P$, the spatial left-right parity}
This again imposes
$\gamma(c_0-c_8)+c_{12}-c_{15}=0$, and the resulting rate function is the same as for the case of detailed balance with $P$ symmetry (see~(\ref{app:sym})).

\subsection*{Symmetry under $CP$}
This imposes the two additional constraints~(\ref{app:cp}) on~(\ref{app:res}), or~(\ref{app:cp2}) 
on~(\ref{app:res2}).
The resulting rate function depends on 4 free parameters.

\subsection{Totally asymmetric dynamics}
For instance, only the flipping of $+-$ into $-+$ is allowed.
Hence $w(-+;\{\s_j\})=0$, or $\w_\alpha=0$.
We thus set the two left sides of~(\ref{app:res}) to zero, since $d_0=d_8=d_{12}=d_{15}=0$, from which it results that
\beq
c_8=0,\qquad \gamma c_0+c_{12}-c_{15}=0,
\eeq
or equivalently
\beq\label{app:tas}
w_1+w_4-w_2-w_3=0,\quad w_2=w_3\e^{-4K}.
\eeq
The resulting rate function reads
\beqa\label{app:ta}
w(+-;\{\s_j\})
=c_0+c_{12}\s_{n-1}+(\gamma c_0+c_{12})\s_{n+2},
\eeqa
which depends on 2 free coefficients.
Imposing the $CP$ symmetry fixes $c_{12}=-\gamma c_0/2$.
The solution found is therefore unique, up to the global time scale $c_0$:
\beq
w(+-;\{\s_j\})
=c_0\left(1-\frac{\gamma}{2}(\s_{n-1}-\s_{n+2})\right).
\eeq

\subsection*{Partially asymmetric dynamics}
A partial asymmetry with uniform bias~\cite{kls} translates into the condition
\beq
\frac{w(-+;\s_{n-1},\s_{n+2})}{w(+-;\s_{n+2},\s_{n-1})}=\frac{1-V}{1+V},
\eeq
where $0\le V\le1$.
This condition yields
\beqa\label{app:pa}
w(+-;\{\s_j\})&=&
c_0+c_{12}\s_{n-1}+(\gamma c_0+c_{12})\s_{n+2},
\nonumber\\
w(-+;\{\s_{j}\})&=&\frac{1-V}{1+V}(c_0+(\gamma c_0+c_{12})\s_{n-1}+c_{12}\s_{n+2}),
\eeqa
which depend on 3 parameters: $c_0, c_{12}$ and $V$.
Eqs.~(\ref{app:tas}) still hold.
The limiting case $V=0$ is included in the solution~(\ref{app:sym}) of the fully symmetric case.
The totally asymmetric limit $V=1$ reproduces the result~(\ref{app:ta}).
Imposing the $CP$ invariance on the rate function again fixes $c_{12}=-\gamma c_0/2$.

%

\end{document}